# A DUAL DIGITAL SIGNAL PROCESSOR VME BOARD FOR INSTRUMENTATION AND CONTROL APPLICATIONS*

H. Dong, R. Flood, C. Hovater, J. Musson
Thomas Jefferson National Accelerator Facility, Newport News, VA

## Abstract

A Dual Digital Signal Processing VME Board was developed for the Continuous Electron Beam Accelerator Facility (CEBAF) Beam Current Monitor (BCM) system at Jefferson Lab. It is a versatile general-purpose digital signal processing board using an open architecture, which allows for adaptation to various applications. The base design uses two independent Texas Instrument (TI) TMS320C6711, which are 900 MFLOPS floating-point digital signal processors (DSP). Applications that require a fixed point DSP can be implemented by replacing the baseline DSP with the pin-for-pin compatible TMS320C6211. The design can be manufactured with a reduced chip set without redesigning the printed circuit board. For example it can be implemented as a single-channel DSP with no analog I/O.

## 1 HARDWARE

### *1.1 Board Digital Signal Processor*

Two Texas Instrument TMS320C6711 floating-point digital signal processors power the board. This processor provides cost effective solutions to high-performance digital signal processing programming challenges with its 900 million floating-point operations per second (MFLOPS) at a clock rate of 150 MHz. As shown in Figure 1, this processor has sixteen 32-bit registers and eight independent functional units that provide floating-/fixed-point arithmetic logic unit (ALU) and multipliers to produce two multiply and accumulate instructions (Macs) for a total of 300 million Macs. In addition, its two-level cache-base architecture facilitates zero wait state program execution. The Level 1 program cache (L1P) is a 4K by 32 bits direct mapped cache and the Level 1 data cache is a 4K by 32 bits 2-way set-associate cache. The 16K by 32 bits Level 2 (L2) memory serves as either memory or cache or both. This flexibility allows the program to reside either inside or outside the chip with no loss of performance for most programs. Finally, with its diverse set of peripherals, this processor board can either operate stand-alone or be expanded to include additional peripherals. This set includes two multi-channel buffered serial ports (McBSPs), two general-purpose timers, a host-port interface (HPI) to allow direct access to all locations, and an external memory interface (EMIF) for expansion. The EMIF supports synchronous dynamic random access memory (SDRAM), synchronous burst dynamic RAM (SBSRAM), and asynchronous peripherals. [1]

### *1.2 Board Peripherals*

The EMIF supports the board peripherals. As shown in Figure 2, these include 256K by 32-bit FLASH electrically erasable programmable read only memory (EEPROM), 4Meg by 32 SDRAM, 64K x 16 dual port RAM, eight digital inputs, eight digital output, a watch dog timer, 12-bit 3us DAC, 12-bit 333ks/s ADC, mezzanine connectors for expansion, and a field programmable gate array (FPGA). The peripherals are mapped to EMIF CE3, CE2, CE1, and CE0 EMIF's memory spaces. McBSP signals are connected to support either global (with other boards) or local (between the DSP) communication. If cost is a concern, these peripherals can be populated as needed.

### *1.3 VME Interface*

In the CEBAF Accelerator, the FPGA is programmed to support VME bus protocols and BCM requirements. The VME protocols supported are 16 bits slave transfer (D16), Even/Odd byte transfer (D08(EO)), Block transfer cycle (BLT), Read modified write cycle (RMW), Address only cycle (ADO), Address pipelining, and Selectable Interrupt Request with D08 and D16 Status/ID. The FPGA can be re-programmed to support different protocols and applications. For example the FPGA can be re-programmed to provide a communication link to other boards and to support multi-processors system.

### *1.4 Expansion*

Dual 160-pin mezzanine connectors provide expansion capability without design modifications. Daughter boards can be built to provide additional functions. This interface conforms to the TI Expansion Daughter Card Interface standard and it can support multi-processors configuration. In the BCM system, this is used to interface to the Digital Down Converter (DDC) receiver board. The DSP configures the DDC and receives in-phase and quadrature (I:Q) beam current intensity data through this interface.

* This work was supported by the U.S. DOE Contract No. DE-AC05-84-ER40150

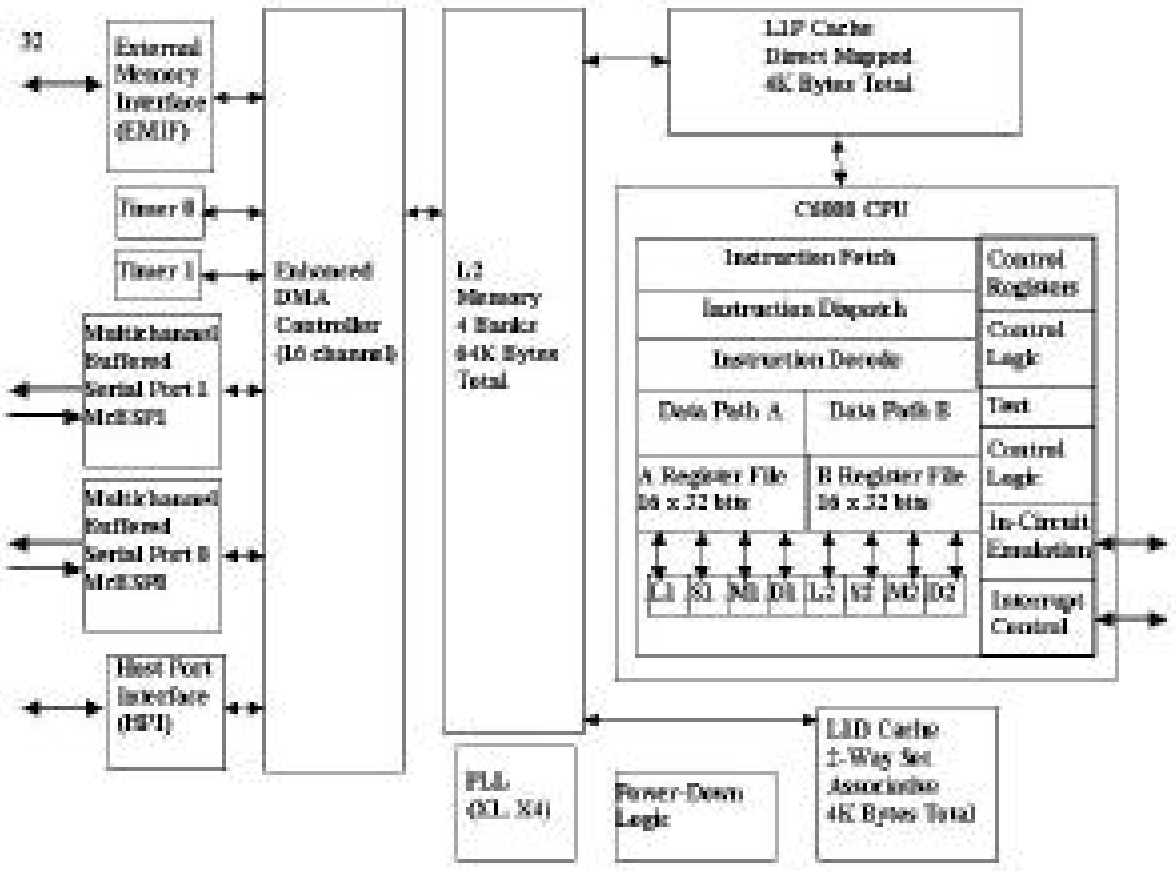

Figure 1 [1]: TMS320C6711 Digital Signal Processors

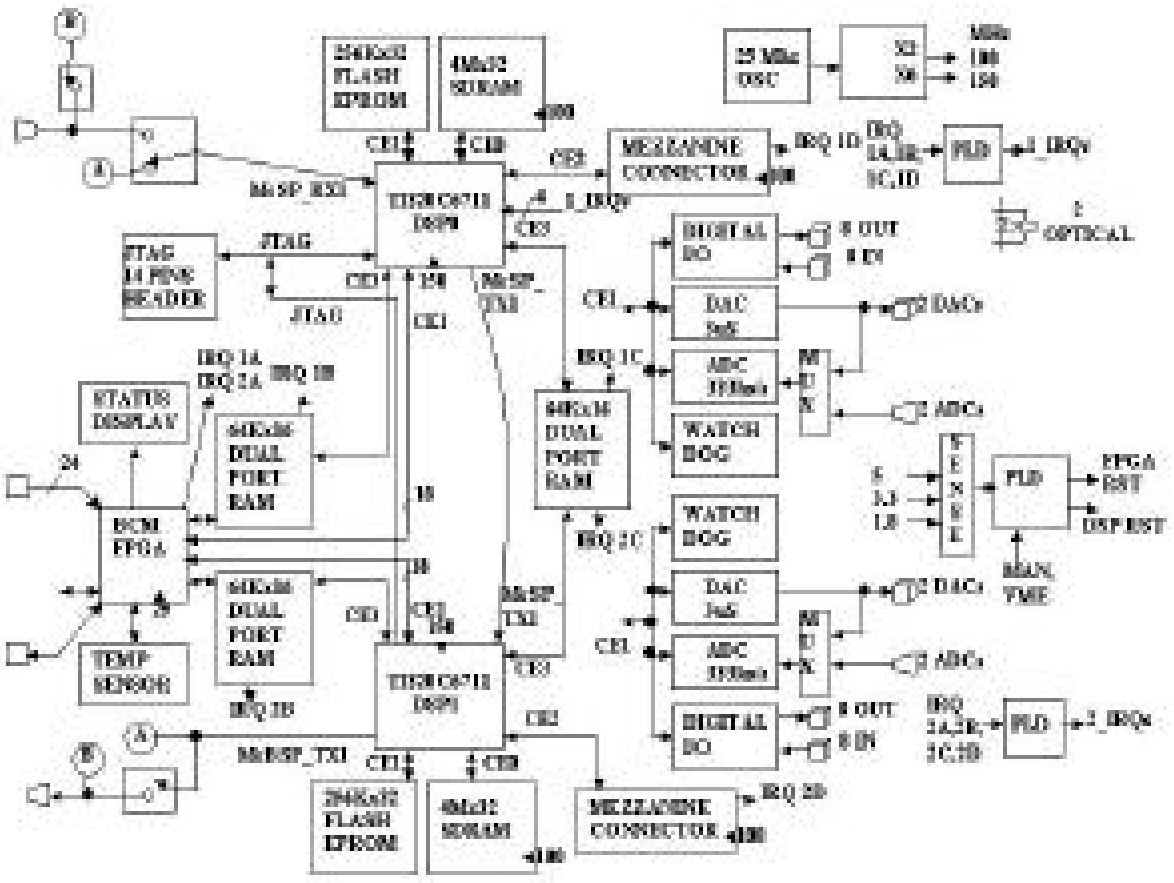

Figure 2: Block Diagram for Dual DSP Board

## 2 SOFTWARE

### 2.1 Program Development Tools

Texas Instrument provides a complete set of development tools to simplify program development. These tools include a C/C++ compiler, an assembly optimizer, a Windows debugger interface for visibility into source code execution, and a loader to load the program to DSP. Besides using these tools for software development, these tools were used to facilitate the testing and debugging of the hardware. These tools are bundled in TI Code Composer Studio software package. Texas Instrument also provides various digital-signal-process (DSP) algorithms to help with DSP applications. The application programs are loaded to the DSP via Spectrum Digital DS510PP_PLUS JTAG emulator.

### *2.2 Jefferson Lab-Developed Programs*

Various programs written at Jefferson Lab facilitate firmware development and hardware checkout. For example, one program stores the application code into the FLASH EEPROM eliminating the need for a separate EEPROM programmer. Another program, which can be included with the application code as power on self-test (POST), tests and verifies various components on the board.

## 2 PERFORMANCE

Various optimized C functions were run to gauge the performance of the TI C6711 DSP. A digital output bit is toggled before and after a function is executed. Table 1 lists the time taken to execute the functions tested.

Table 1: DSP Performance Table

| Functions | Time to Execute in ns |
|---|---|
| Asynchronous R/W | 120 |
| For loop<br>for (;;); | 130 |
| Multiply<br>C = A * A | 91 |
| Add<br>C = A + A | 91 |
| Square Root<br>C = sqrt(A) | 2820 |
| Magnitude<br>C = sqrt (A*A + B*B) | 3070 |
| 8, 16, 32 tap FIR | 5000, 9600, 17600 |

## 4 BCM IMPLEMENTATION

### *4.1 Overview*

In the BCM system, the board is currently being programmed to satisfy the requirement of a Beam Loss Accounting System. The program has to perform both housekeeping functions and process beam current data. House keeping functions include processing VME commands, output beam current data to DAC for monitoring, and configuring the DDC board. Beam processing includes filtering I/Q data coming from the DDC and calculating the magnitude.

### 4.2 Adaptive Filtering

Adaptive Filtering algorithms are being developed to filter low S/N beam current. Adaptive filtering is used to filter out narrowband noise as well as remove discrete sinusoid components. The algorithm varies filter parameters in real time to achieve certain criteria (i.e. maximize S/N). Two algorithms are being evaluated: The Predictor Topology and the Identification Topology shown in Figure 3 incorporate a finite impulse response (FIR) filter or an infinite impulse response (IIR) filter with a least-means squares (LMS) algorithm to vary the coefficients. [2][3]

### 4.3 DSP Adaptation

Adaptive Filtering algorithm adapts seamlessly with the board. While one DSP executes the FIR or IIR algorithm, the other DSP performs the LMS routine. The data and filter's coefficients, which are stored in the dual port memory, are accessed by the second DSP with no overhead. The algorithm that requires less time to execute can be assigned to the DSP that manages the overhead functions.

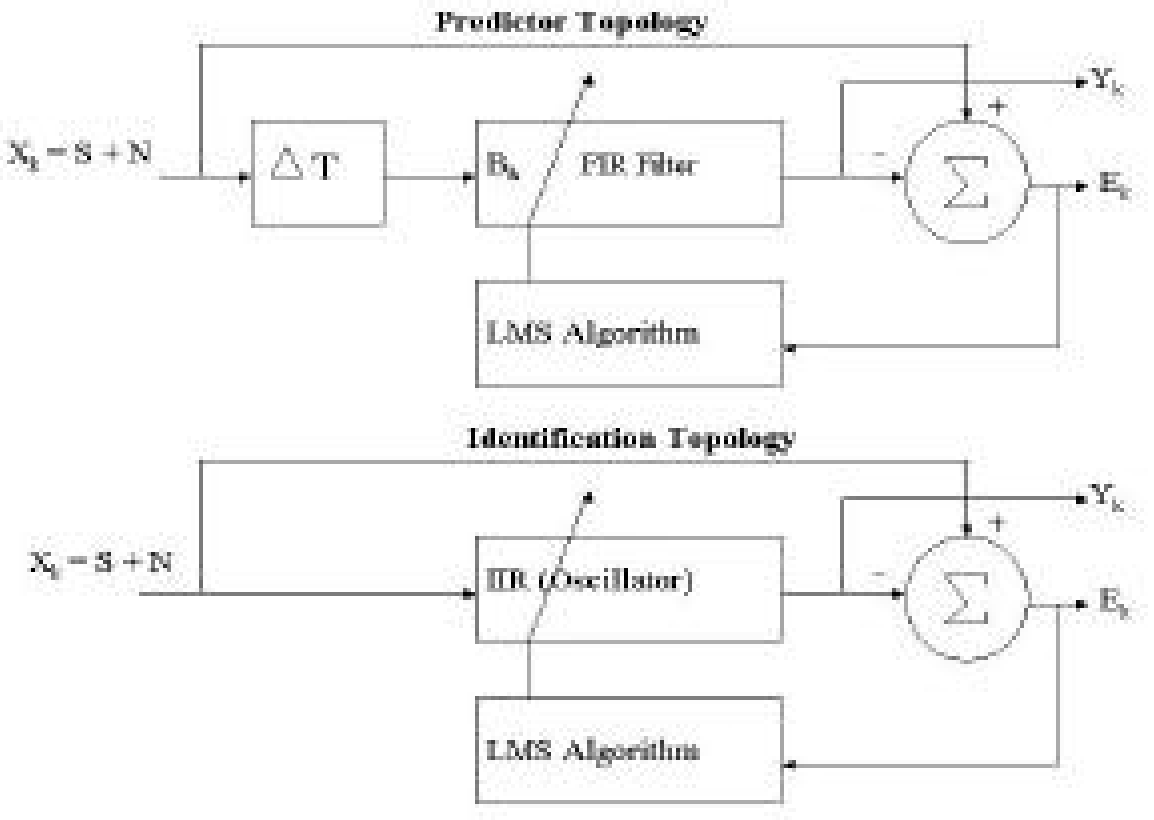

Figure 3 [2]: (a) Adaptive Prediction Topology where $X_k$ is input signal plus noise, while predicted and error signals are $Y_k$ and $e_k$, respectively. FIR coefficients are denoted by $b_k$. (b) Adaptive Identification topology.

## 5 OTHER IMPLEMENTATIONS

Beam Positioning Monitor (BPM) systems can take advantage to the computational power of this DSP board [4]. In this system, one DSP calculates the X position and the other DSP calculates the Y position. The results can be output to the on-board DAC as well as the accelerator control system. The DSP can also provide additional filtering algorithm to improve signal to noise.

Low Level Radio Frequency (LLRF) system for 12-GEV-upgrade being developed at JLAB incorporates the board in the design. As shown in Figure 4, the DSP provides gain control, adaptive control, resonance, loop phase, quench interlock, system calibration, and data acquisition.

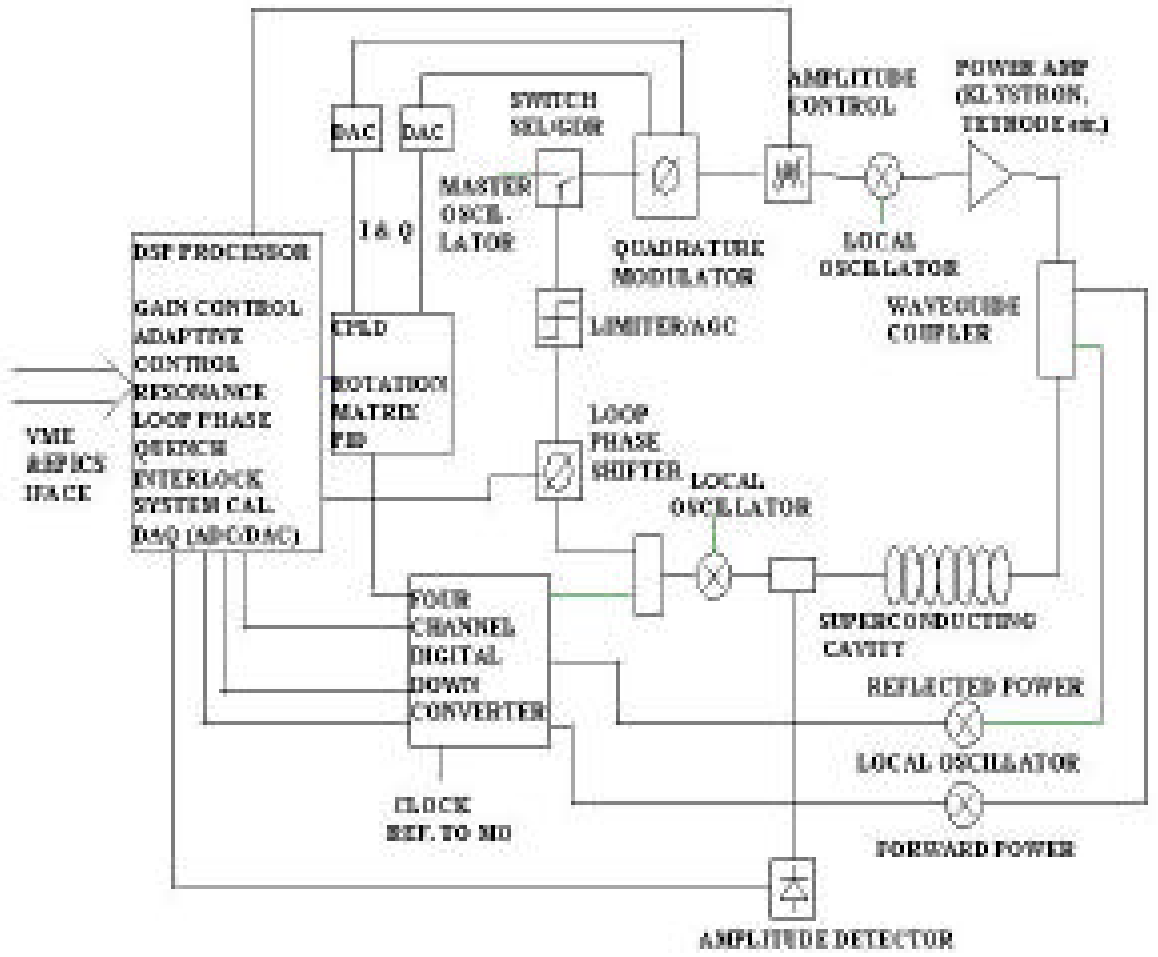

Figure 4: Conceptual LLRF System

## 6 CONCLUSION

The open architecture coupled with the processing power of the TMS320C6711 allows this board to be used for other applications besides the BCM. The board is supported by a complete set of development tools to simplify program development and hardware testing. Finally, the board can be manufactured with a reduce chip set for cost saving.